\documentstyle[12pt]{article}
\begin{document}

\hfill PUPT-1660

\hfill hep-th/9611196

\vspace{1.5in}

\begin{center}

{\large\bf Boundary Superpotentials }

\vspace{1in}

Eric Sharpe \\
Physics Department \\
Princeton University \\
Princeton, NJ  08544 \\
{\tt ersharpe@puhep1.princeton.edu }

\vspace{0.75in}

\end{center}

In this paper we work out explicit lagrangians describing superpotential
coupling to the boundary of a 5D orientifold, as relevant to a number of
quasi-realistic models of nature.  We also make a number of general
comments on orientifold compactifications of M theory.

November 1996

\newpage

{\bf 1.  Introduction }

Historically heterotic string theory has been the starting point for
phenomenologically promising string vacua, with N=1 supersymmetry in 4D.  
Although recent advances in
the understanding of string duality have put other theories on an equal
footing, it is still gives the most realistic concrete string
vacua and is a convenient starting point for many studies.

The heterotic string theory compactified to four dimensions has
a variety of possible strong coupling limits.  One of them,
which is inherited from the strong coupling limit in ten dimensions,
involves the appearance of a fifth large dimension.  In this particular
limit, the heterotic string theory compactified to $R^4$ (on 
a compact six-manifold $X$) is equivalent to 
eleven-dimensional M theory compactified (on the same $X$)
to $R^4\times S^1/Z_2$.
Taking account of actual numerical values of coupling constants,
as explained in detail in    \cite{bd,bdII}, it appears that the 
$S^1/Z_2$ can be much larger than $X$, so that in this regime,
which actually has some possible phenomenological virtues, 
the universe is quasi-five-dimensional.

The purpose of this paper is to study physics in this
quasi-five-dimensional regime.  We will in fact focus on the
question of the superpotential.  There is no superpotential
in supersymmetric theories in five dimensions, but superpotentials
can definitely be present in $N=1$ supersymmetric dynamics in four
dimensions, and play a crucial role.  How can a superpotential
be present in a world that macroscopically looks like $R^4\times
S^1/Z_2$?  The answer must be that as the superpotential is
impossible in bulk -- where the world looks five-dimensional --
it must appear as a boundary interaction.  That explains the title
and theme of this paper: we will consider supersymmetric field
theories on $R^4\times S^1/Z_2$ and search for boundary interactions
that generate an effective superpotential for the massless
four-dimensional modes.

The paper is organized as follows.
In section two, we describe several classes of models to which
the discussion applies, of which the strong coupling limit of the
heterotic string is actually only one.
In sections three and four we describe in rigid supersymmetry
the boundary couplings that gives superpotentials (the extension
to supergravity will be discussed elsewhere \cite{me}).
We begin in section three with just hyper plets in the 5D bulk,
and demonstrate how to couple the boundary chiral plets descending
from hyper plets to a superpotential.  We also discuss coupling
boundary-confined fields to the same superpotential.
In section four 
we generalize to theories with both vector plets and hyper plets
in the bulk.  In this case there are two physically distinct possible
orientifolds, and we discuss superpotential couplings for both cases.
In section five, we discuss anomalies and the compactification
of the Green-Schwarz mechanism.  In section six we conclude, with 
a discussion of some novel supersymmetry-breaking mechanisms possible
in these orientifold theories, and some novel insights in
nonperturbative effects in heterotic string theory.

{\bf 2.  Examples of orientifold compactifications}

Actually, the discussion here has certain applications
that go beyond the motivation that was stated above.
Our framework is relevant to any $N=1$ supersymmetric model
on a space-time that looks macroscopically like $R^4\times S^1/Z_2$.
The strong coupling limit of the heterotic string is an important
example, but there are several other examples.

To describe some of these systematically, start with
M theory on $Y=R^4\times S^1\times X$, where $X$ is a 
Calabi-Yau six-manifold.  To obtain an $N=1$ model
in four dimensions, we divide $Y$ by a $Z_2$ symmetry $\tau$
that acts trivially on $R^4$, acts as $-1$ on $S^1$, and
acts as some involution $\tau'$ (or $Z_2$ symmetry) of $X$.
Supersymmetry will hold under several possible conditions on
$\tau'$:

(1) If $\tau'$ acts trivially on $X$, we get M theory on
$R^4\times S^1/Z_2\times X$, which is the strong coupling limit
of the heterotic string, discussed above as a motivating example.

(2) If $\tau'$ acts on $X$ in a way that preserves the complex
structure and the holomorphic three-form, we get a more
general supersymmetric orientifold.  The fixed points of such
a $\tau'$ on $X$ will be curves\footnote{Consider the normal bundle to
the fixed-point set.  As $\tau'$ preserves both the complex structure
and the holomorphic three-form, it must be the case that the normal
bundle has even rank, so as $X$ has complex dimension three, the fixed
point set must have dimension one or three.  The latter corresponds to
case (1).}  
of $A_1$ singularities,
so the fixed points of $\tau$ on $Y$ will look locally like
$R^5/Z_2$.  Such a model
may also have five-branes wrapped around two-cycles in 
$(S^1\times X)/Z_2$ in analogy to \cite{fivebrane}. 

(3) If $\tau'$ is an anti-holomorphic isometry that reverses
the complex structure of $X$, then $(S^1\times X)/Z_2$ is a
seven-dimensional orbifold whose structure group -- an extension
of $SU(3)$ by $Z_2$ -- is a subgroup of $G_2$.  Such an orbifold
can plausibly be deformed to a smooth manifold of $G_2$ holonomy
-- as in the work of Joyce \cite{joyce} -- and such orbifolds would
appear to make sense in M theory even if they cannot be so deformed.

Let us discuss the three cases in more detail.

(1)  First, recall the compactification of M theory on $X$ \cite{anton}.
In 5D we get $h_{1,1}$ vectors
($h_{1,1}-1$ vector plets),
and $h_{2,1}+1$ hyper plets.  The vectors all descend from the 11D
3-form potential.  There are $h_{2,1}$ hyper plets each consisting of a
complex boson obtained from zero modes of the 11D metric along complex
structure deformations, and a pair of real bosons obtained from the 11D
3-form on elements of $H^{2,1}$ and $H^{1,2}$.  In addition, there is
one hyper plet consisting of the volume of $X$, the real scalar
dual to a 4D 3-form potential, and two scalars obtained from the 11D
3-form on the holomorphic and antiholomorphic 3-forms of $X$.

Under the orientifold action
\cite{pwI,pwII,pIII}, the 11D 3-form potential is odd, so we can
immediately read off the fact that each of the $h_{1,1}$ 5D vectors is
odd under the orientifold, and half of each of the $h_{2,1}+1$ hyper
plets are projected out by the orientifold, yielding a total of
$h_{1,1}+h_{2,1}+1$ chiral plets on the boundary.  This yields a
new derivation of one of the old spacetime superpotential
nonrenormalization theorems \cite{xw}, as will be discussed in more
detail later.

On each boundary of the 11D theory, one gets an $E_{8}$ vector.
In compactification, a gauge sheaf\footnote{Recall
from \cite{dgm} that even after resolving
the base space of a
heterotic (0,2) compactification, it still may not be possible to
resolve the singularities of the ``bundle," so we shall use more nearly
correct terminology and refrain from calling the gauge sheaves
``bundles."} is embedded in each $E_{8}$,
and the resulting matter
spectrum is well-known \cite{dg}.  Matter fields are in one-to-one
correspondence with sheaf cohomology groups, with Serre duality
exchanging particles and antiparticles.\footnote{
Recall from \cite{dk}, that at
least in certain (non-geometric) heterotic (0,2) compactifications,
moduli associated with bundle deformations and moduli associated with
complex moduli of the base are in some sense intermingled.
Here we see that in M
theory, these moduli are clearly distinguished:  complex structure and
Kahler
moduli of the Calabi-Yau 3-fold propagate in the 11D bulk, whereas
moduli associated with bundle deformations (corresponding to elements of
$H^{m}(X, \mbox{ End } V)$) only propagate on the
boundary.  It is also true that in F theory compactifications \cite{mv},
the base space moduli and the bundle moduli are easily distinguished.
}

(2) Examples of this sort in six dimensions are well-documented in the
literature.  For example, in \cite{sen} Sen considers M theory on $(K3
\times S^{1})/Z_{2}$, with $\tau'$ acting on $K3$ so as to preserve the
complex structure and holomorphic two-form.  In the compactification of
M theory to 7D on $K3$, we get 7D vector plets corresponding to elements
of $H^{2}$; $\tau'$ leaves 14 elements of $H^{2}$ invariant and flips
the sign of the other 8.  As a result, 14 of the 7D vector plets are
projected to 6D hyper plets on the boundary, and 8 are projected to 6D
vector plets on the boundary.  
In addition, Sen assigns magnetic three-form charge
$-1/2$ to each of the fixed points of $\tau'$, and notes that half of
the fixed points of $\tau'$ have five-branes, essentially just as was
done in \cite{fivebrane}.  

By fibering $K3$ over a $P^{1}$, we can
construct ($K3$-fibration) Calabi-Yau 3-folds with an orbifold symmetry
$\tau'$ of the desired form.  As above, if $\tau'$ leaves an element of
$H^{1,1}$ invariant, then the corresponding 5D vector plet will project
to a chiral plet on the boundary; if $\tau'$ flips the sign of an
element of $H^{1,1}$, then the corresponding 5D vector plet will project
to a vector plet on the boundary.  Note that symmetries which act on
elements of $H^{1,1}$ are only present on subsets of Kahler moduli
space, just as symmetries which act on elements of $H^{2,1}$ are only
present on subsets of complex structure moduli space; this fact will be 
reflected in constraints on intersection
numbers,
as will be discussed later in this paper.

Here is a related example of case (2), 
following \cite{se,kmp}.
Recall the action of elementary transformations on 3-folds,
which act on elements $D$ of $H^{2}$ as
\begin{displaymath}
D \: \rightarrow \: D \: + \: (D \cdot C) \, E 
\end{displaymath}
where $E$ is a divisor swept out by a family of holomorphic 2-spheres
$C$.  As $E \cdot C \: = \: -2$, this acts as a reflection on $E$. 
(For additional mathematical details see \cite{elemtrans}.)  As is
well-known, at the fixed point of a symmetry that acts on the
Teichmuller space, one gets an enhanced symmetry -- in this case, when 
the divisor $E$ collapses to a curve.  For example, the Calabi-Yau
$P^{4}_{1,1,2,2,2}[8]$ has a genus 3 curve of $A_{1}$ singularities at
$z_{1} \: = \: z_{2} \: = \: 0$.  At generic points in Kahler moduli
space this curve is blown up into a divisor.  As is well-known from
mirror symmetry \cite{candelas} the Kahler moduli space of this 3-fold
has a $Z_{2}$ quotient singularity, precisely as expected at 
the enhanced symmetry point. 
Returning to our
orientifold, if we orientifold M theory by the action defining the $E_{8} 
\times
E_{8}$ heterotic string and simultaneously orbifold the Calabi-Yau by
this enhanced symmetry, then we not only will freeze the Calabi-Yau
Kahler moduli at a singular point (yielding enhanced gauge symmetry in
5D) but now some of the 5D vector plets will be projected to boundary
vectors (whereas in the standard heterotic orientifold, all 5D
vectors are projected to chiral plets on the boundary).

(3)  As $\tau'$ is an antiholomorphic isometry, some elements of $H^{1,1}$
are odd under the involution, so corresponding 5D vector plets will
be even and will project to vectors on the
boundary.  
If any elements of $H^{1,1}$ are even under $\tau'$, then the
corresponding 5D vector plets will project to chiral plets on the
boundary, as in the previous cases.
In addition, there will be boundary-confined chiral plets reflecting modes
that in perturbative string theory would be associated with twisted
sectors.   

{\bf 3.  Pure hyper plet action in bulk}

To warm up, we will first consider the case that the bulk theory contains
only hyper plets, no vectors.
The bulk 5D N=2 rigid matter action is

\begin{displaymath}
S \: \: = \: \: \int_{M} [ \: - \frac{1}{2} \eta^{\mu \nu} g_{x' y'} 
\partial_{\mu}
\sigma^{x'} \partial_{\nu} \sigma^{y'} \: - \:
\frac{1}{4} \overline{\lambda}^{a'} \Gamma^{\mu} D_{\mu} 
\lambda_{a'} \: + \: 4 \mbox{ fermi } ]
\end{displaymath}

where
\begin{eqnarray*}
D_{\mu} \lambda_{a'} & = & \partial_{\mu} \lambda_{a'} \: + \:
(\partial_{\mu} \sigma^{x'}) \, \omega_{x' a' b'} \, \lambda^{b'} \\
\partial_{y'} f_{x'}^{i a'} & = & - \omega_{y' \: b'}^{ \: a'}
f_{x'}^{i b'} \: + \: \Gamma^{z'}_{x' y'} f_{z'}^{i a'} \\
\delta \sigma^{x'} & = & \frac{i}{2} \, f_{i a'}^{x'} \,
\overline{\epsilon}^{i} \lambda^{a'} \\
\delta \lambda^{a'} & = & -i \, f_{x'}^{i a'} \, \Gamma^{\mu}
\partial_{\mu} \sigma^{x'} \, \epsilon_{i} 
\end{eqnarray*}

Conventions closely follow those of \cite{Sierra:matter,sugrav5Dmain,nonabel},
and are also outlined in an appendix.

$x'$ labels coordinates $\sigma$ on the hyperKahler manifold 

$a'$ is an index in the fundamental of Sp(n); it labels the fermions $\lambda$
of
the n hyperplets

$i$ is an index in the fundamental of Sp(1)

The orientifold symmetry is as follows.
Under $x^{\mu} \: \rightarrow \: + \, x^{\mu}$, for $\mu \: < \: 5$,
and $x^{5} \: \rightarrow \: - \, x^{5}$, the fermions transform as: 
\begin{eqnarray*}
\lambda^{a'} & \rightarrow & + i \, \Gamma_{5} \, \lambda_{a'} \\
\lambda_{a'} & \rightarrow & - i \, \Gamma_{5} \, \lambda^{a'} 
\end{eqnarray*}

We assume the hyperKahler manifold
has an orientifold symmetry which preserves one of the complex
structures (but not the others), so that in a coordinate patch containing
the boundary of the orientifolded hyperKahler manifold,
half the coordinates are invariant (call these $\sigma^{x1}$ and
$\sigma^{x2}$) and the other half get a sign flip (call these
$\sigma^{x3}$ and $\sigma^{x4}$).  In other words, 
in a basis adapted to the complex structure preserved by the
orientifold, in each hyper plet one chiral plet is even under the 
involution and one is odd. 
[After coupling to supergravity \cite{me}, we will assume that the quaternionic
manifold of scalars has such an involution, whose fixed point set is
also Kahler.]

Under the orientifold the ``veirbein" f transforms as
\begin{displaymath}
f_{x'}^{i a'} \: \rightarrow \: f_{x' i a'}
\end{displaymath}

The supersymmetries that commute with the orientifold are given by
\begin{eqnarray*}
\epsilon_{i} \: & = & \: + i \, \Gamma_{5} \, \epsilon^{i} \\
\epsilon^{i} \: & = & \: - i \, \Gamma_{5} \, \epsilon_{i}
\end{eqnarray*}

The complete action describing the coupling of an N=1 superpotential $W$ on the boundary is
\begin{displaymath}
S_{bulk} \: \: = \: \: \int_{M} [ \: - \frac{1}{2} \eta^{\mu \nu} g_{x' y'} 
\partial_{\mu}
\sigma^{x'} \partial_{\nu} \sigma^{y'} \: - \:
\frac{1}{4} \overline{\lambda}^{a'} \Gamma^{\mu} D_{\mu}
\lambda_{a'} \: + \: 4 \mbox{ fermi } ] \: \:
\end{displaymath}
\begin{displaymath} 
S_{boundary} \: \: = \: \: \int_{\partial M} [ \:  g^{ x \overline{y} } ( 
\partial_{x} W )
\, 
( \partial_{\overline{y}} \overline{W} ) \, \delta (0) \: - \:
\frac{1}{8} \, ( D_{x} D_{y} W ) \, \overline{\chi}^{x} \chi^{y} \: + \:
\frac{1}{8} \, ( D_{\overline{x}} D_{\overline{y}} \overline{W} ) \, 
\overline{\chi}^{\overline{x}} \chi^{\overline{y}} \: + \: 4 \mbox{
fermi } ]
\end{displaymath}
\vspace{0.15in}
\begin{displaymath}
S \: = \: S_{bulk} \: + \: S_{boundary} 
\end{displaymath}

where, in the bulk theory,
\begin{eqnarray*}
D_{x} D_{y} W \: & = & \partial_{x} \partial_{y} W \: - \: \Gamma^{z}_{x
y} \, \partial_{z} W \\
D_{\mu} \lambda_{a'} & = & \partial_{\mu} \lambda_{a'} \: + \:
(\partial_{\mu} \sigma^{x'}) \, \omega_{x' a' b'} \, \lambda^{b'} \\
\delta \sigma^{x'} & = & \frac{i}{2} \, f_{i a'}^{x'} \,
\overline{\epsilon}^{i} \lambda^{a'} \\
\delta \lambda^{a'} & = & -i \, f_{x'}^{i a'} \, \Gamma^{\mu}
\partial_{\mu} \sigma^{x'} \, \epsilon_{i} \: \: + \: \: \delta( x_{5} )
F^{i a'} \epsilon_{i} \\
F^{1 A } & = & + ( \partial_{x} W ) \, f_{1 A}^{x}
\: + \: ( \partial_{\overline{x}} \overline{W} ) \, f_{1
A}^{\overline{x}} \\
F^{ 2 A } & = & - (\partial_{x} W) \, f_{2
A }^{x} \: - \: (\partial_{\overline{x}} \overline{W}) \,
f_{2 A}^{\overline{x}} \\
F^{i a'} & = & - \: F_{i a'}
\end{eqnarray*}

Factors of $\delta(0)$ in this paper appear
for essentially the reasons discussed by Horava and Witten in
\cite{pwI,pwII,pIII}.

We have defined the boundary fields
\begin{eqnarray*}
A^{x} & = & \sigma^{x1} \: + \: i \, \sigma^{x2} \\
\chi^{A} & = & \lambda^{A} \: - \: i \,
\lambda^{A'} \\
\chi^{x} & = & 2 \, [ \, f_{1 A}^{x1} \, + \, i f_{1
A}^{x2} \, ] \, \chi^{A} \\
 & = & 2 \, f_{1 A}^{x} \chi^{A} \\
\epsilon & = & \epsilon^{1} \: + \: i \, \epsilon^{2} \\
\epsilon^{*} & = & \epsilon^{1} \: - \: i \, \epsilon^{2}
\end{eqnarray*}

In other words, $A^{x}$ is the complex boson that is even under the involution.

As is typical in 4D N=1 supersymmetry, the superpotential $W$ is
holomorphic in the chiral superfields.

Note that in these conventions, $\epsilon$ and $\chi^{x}$ both have 
positive chirality.

Note that $\epsilon^{*}$ is not the complex conjugate of $\epsilon$,
but in fact is merely another linear combination of spinors,
of opposite four-dimensional chirality, convenient for our purposes here.

By restricting the bulk supersymmetry transformations to the boundary we find
\begin{eqnarray*}
\delta A^{x} & = & \frac{i}{4} \, \overline{\epsilon} \chi^{x} \\
\delta A^{\overline{x}} & = & \frac{i}{4} \, \overline{\epsilon}^{*}
\chi^{\overline{x}} \\
\delta \chi^{x} & = & - \, \Gamma^{\mu} ( \partial_{\mu} A^{x} ) \,
\epsilon^{*} \: + \: i \delta (0) \, g^{ x \overline{y} } \, (
\partial_{\overline{y}} \overline{W} ) \, \epsilon \: + \: 3 \mbox{ fermi} \\
\delta \chi^{\overline{x}} & = & + \, \Gamma^{\mu} ( \partial_{\mu}
A^{\overline{x}} ) \, \epsilon \: - \: i \delta (0) \,
g^{\overline{x} y } \, ( \partial_{y} W ) \, \epsilon^{*} \: + \: 3
\mbox{ fermi}
\end{eqnarray*}

Recall that $x'$ in the bulk theory denotes a coordinate on the
hyperKahler space; here, $x$ and $\overline{x}$ denote holomorphic
and antiholomorphic coordinates on the Kahler manifold.
Also, recall that $a'$ was in the fundamental of Sp(n);
on the boundary, we've split $a' \: = \: 1 ... 2n$ into
$A \: = \: 1 \ldots n$ and $A' \: = \: n+1 \ldots 2n$.

There is a constraint on the coordinates $\sigma^{x1}$,
$\sigma^{x2}$:
\begin{eqnarray*}
f_{1 A}^{x1} & = & + \: f_{2 A}^{x2} \\
f_{1 A}^{x2} & = & - \: f_{2 A}^{x1}
\end{eqnarray*}

This constraint insures that in the coordinates $A^{x} \: = \: \sigma^{x1}
\: + \: i \, \sigma^{x2}$ on the boundary of the hyperKahler manifold, the 
metric
is hermitian, which need not be true for an arbitrary set of
coordinates.  Note that this constraint is also necessary in order to
recover a well-defined chiral plet. 

After coupling to supergravity \cite{me} we will recover analogous
results.  For example, the purely bosonic terms in the superpotential
coupling will also contain a $\delta (0)$ factor, as here.

In addition to coupling chiral plets $( \, A^{x} \, , \, \chi^{x} \, )$
descending from bulk hyper plets, we can also couple chiral plets 
$( \, B^{x} \, , \, \psi^{x} \, )$ that are confined to the boundary
to the same superpotential.  The complete lagrangian is
\begin{eqnarray*}
S & = \: \int_{M} \, [ & - \frac{1}{2} g_{x' y'} \, \partial_{\mu}
\sigma^{x'} \, \partial^{\mu} \sigma^{y'} \: - \: \frac{1}{4}
\overline{\lambda}^{a'} \Gamma^{\mu} D_{\mu} \lambda_{a'} \, ] \\
 & + \: \int_{\partial M} \, [ & - g_{x \overline{y}} \, \partial_{\mu}
B^{x} \, \partial^{\mu} B^{\overline{y}} \: - \: \frac{i}{4} g_{x
\overline{y}} \, \overline{\psi}^{x} \Gamma^{\mu} D_{\mu}
\psi^{\overline{y}} \, ] \\
& + \: \int_{\partial M} \, [ & g^{x \overline{y}} \, \frac{\partial
W}{\partial B^{x}} \frac{\partial \overline{W}}{\partial
B^{\overline{y}}} \: + \: \delta (0) \, g^{x \overline{y}} \frac{\partial
W}{\partial A^{x}} \frac{\partial \overline{W}}{\partial
A^{\overline{y}}} \\
 & & - \: \frac{1}{8} \frac{ D^{2} W}{D A^{x} D A^{y}}
\overline{\chi}^{x} \chi^{y} \: + \: \frac{1}{8} \frac{ D^{2}
\overline{W}}{ D A^{\overline{x}} D A^{\overline{y}} }
\overline{\chi}^{\overline{x}} \chi^{\overline{y}} \\
 & & - \: \frac{1}{8} \frac{ D^{2} W }{D B^{x} D B^{y}}
\overline{\psi}^{x} \psi^{y} \: + \: \frac{1}{8} \frac{D^{2}
\overline{W}}{D B^{\overline{x}} D B^{\overline{y}} }
\overline{\psi}^{\overline{x}} \psi^{\overline{y}} \\
 & & - \: \frac{1}{4} \frac{D^{2} W}{D A^{x} D B^{y}}
\overline{\chi}^{x} \psi^{y} \: + \: \frac{1}{4} \frac{ D^{2}
\overline{W}}{D A^{\overline{x}} D B^{\overline{y}}}
\overline{\chi}^{\overline{x}} \psi^{\overline{y}} \: + \: 4 \mbox{
fermi } ]
\end{eqnarray*}

where
\begin{eqnarray*}
D_{\mu} \psi^{\overline{y}} & = & \partial_{\mu} \psi^{\overline{y}} \:
+ \: (\partial_{\mu} B^{\overline{z}})
\, \Gamma^{\overline{y}}_{\overline{z} \overline{w}} \, \psi^{\overline{w}}
\\
D_{x} D_{y} W & = & \partial_{x} \partial_{y} W \: - \: \Gamma^{z}_{x y}
\partial_{z} W
\end{eqnarray*}

We have assumed the Kahler metric $g_{x \overline{y}}$ factorizes
between the chiral plets descending from bulk hypers and the chiral
plets confined to the boundary.  
(More generally, we would expect that the space of all chiral plets on
the boundary has the structure of the total space of a bundle over the
space of chiral plets descending from bulk $(A^{x})$, as we should be
able to consistently forget 
the boundary-confined chiral plets $B^{x}$.) 
Supersymmetry transformations for the
$( \, A^{x} \, , \, \chi^{x} \, )$ chiral plet are as before;
supersymmetry transformations for the chiral plets confined to the
boundary $( \, B^{x} \, , \, \psi^{x} \, )$ are
\begin{eqnarray*}
\delta B^{x} & = & \frac{i}{4} \overline{\epsilon} \psi^{x} \\
\delta B^{\overline{x}} & = & \frac{i}{4} \overline{\epsilon}^{*}
\psi^{\overline{x}} \\
\delta \psi^{x} & = & - \, \Gamma^{\mu} (\partial_{\mu} B^{x})
\, \epsilon^{*} \: + \: i \, g^{x \overline{y}} \, \left( \frac{ \partial \overline{W}
}{\partial B^{\overline{y}} } \right) \,  \epsilon \: + \: 3 \mbox{
fermi } \\
\delta \psi^{\overline{x}} & = & + \, \Gamma^{\mu} (\partial_{\mu}
B^{\overline{x}} ) \, \epsilon \: - \: i \, g^{\overline{x} y} \, \left(
\frac{ \partial
W}{\partial B^{y}} \right) \,  \epsilon^{*} \: + \: 3 \mbox{ fermi }
\end{eqnarray*}

Note that these supersymmetry transformations are identical to those of
the $( \, A^{x} \, , \, \chi^{x} \, )$ multiplet, except that
$\delta(0)$ factors have been dropped.

Just to re-emphasize the point, we have, above, explicitly coupled both
chiral plets descending from bulk and chiral plets confined to the
boundary to the same superpotential.

How should the $\delta(0)$ factor in the bosonic potential
terms be interpreted?   
As mentioned earlier, in general factors of $\delta(0)$ in this paper
occur for essentially the same reasons discussed by Horava and Witten in
\cite{pwI,pwII,pIII}.  More specifically, in the lagrangian above 
the bosonic potential term with the $\delta(0)$ factor corresponds to
superpotential couplings among chiral plets descending from bulk hypers
-- which, in a standard heterotic string compactification, 
correspond to complex and Kahler moduli.  The fact that these
superpotential couplings have a bosonic potential weighted by
$\delta(0)$ suggests that they are suppressed.
Indeed, for heterotic
compactifications with the standard embedding in the gauge sheaf, 
complex and Kahler deformations ((2,2) moduli) are moduli of the theory,
as was shown by Dixon in \cite{dixontrieste}.  A generalization of this
lagrangian to supergravity low-energy effective actions \cite{me} would
naively appear to imply a more general result than Dixon's, but a word of
caution is in order.  As will be discussed later, worldsheet instantons
can not be described in the present framework, 
and as is well-known (0,2) heterotic string compactifications are
sometimes destabilized by worldsheet instantons 
\cite{xw}.  

In passing, it should also be noted that remaining terms in the superpotential
which do not have $\delta(0)$ factors in the bosonic potential, are
often necessarily nonvanishing.
For example, in a heterotic string compactification to 4D N=1,
recall from \cite{stromyuk,dg} that if we embed a gauge sheaf
$V$ of rank 3 in an $E_{8}$, then the superpotential on the
corresponding boundary will contain
$27^{3}$ and $\overline{27}^{3}$ couplings of $E_{6}$ 
proportional to $H^{1}(X, V)^{3}$ and $H^{1}(X,
\wedge^{2} V)^{3}$, at sigma model tree level.  These couplings are
couplings of chiral plets confined to the boundary -- $( \, B^{x} \, ,
\, \psi^{x} \, )$ -- and so the bosonic potential for these plets has no
$\delta(0)$ factor.    

Finally, we would also like to mention that the lagrangian above can be
generalized to include coupling the boundary-confined
chiral plets to boundary-confined vectors, in a very straightforward
fashion.
As these fields are all defined only on the boundary, we can do standard
4D field theory, and recover standard results 
such as the Konishi anomaly \cite{konishi}:
\begin{displaymath}
\{ \, \overline{Q}^{*} \, , \, tr \psi^{\overline{x}} \, B^{y} \, \} \:
= \: -i \, g^{\overline{x} z} \, tr \, ( \,
\partial_{z} W \, B^{y} \, ) \: + \: c \, 
\delta^{\overline{x} y} \, tr \overline{\lambda} \lambda
\end{displaymath}
 
where c is a constant we've not been careful about, $\lambda$ is the
gaugino superpartner of the boundary-confined vector, and $Q^{*}$ is one
of the supersymmetry generators:
\begin{displaymath}
\delta_{\mbox{susy}} \: = \: \overline{\epsilon} Q \: + \:
\overline{\epsilon}^{*} Q^{*} 
\end{displaymath}

{\bf 4.  Vectors in bulk}

The bulk 5D action for nonabelian vectors is
\begin{eqnarray*}
S_{bulk} & =  \int_{M} [ \, &  - \frac{1}{2} g_{xy} D_{\mu}
\phi^{x} D^{\mu} \phi^{y} \: - \: \frac{1}{4} a_{IJ} F^{I}_{\mu
\nu} F^{J \mu \nu} \: - \: \frac{1}{2} \overline{\lambda}^{ia}
\Gamma^{\mu} D_{\mu} \lambda_{i}^{a}  \\
 & & - \: \frac{1}{2 \sqrt{6}} C_{IJK} h^{J}_{a} h^{K}_{b} (
\overline{\lambda}^{i a} \Gamma^{\mu \nu} \lambda_{i}^{b} ) F^{I}_{\mu
\nu}   - \: i \, ( \overline{\lambda}^{ia}
\lambda_{i}^{b} ) \, K^{x}_{I} f^{a}_{x} h^{I}_{b} \\
 & & + \: \frac{1}{ 6 \sqrt{6}} \epsilon^{\mu \nu \rho \sigma \lambda}
\, C_{IJK} \, ( (d A)^{I}_{\mu \nu} (d A)^{J}_{\rho \sigma}
A^{K}_{\lambda} \:
+ \frac{3}{2} (d A)^{I}_{\mu \nu} A^{J}_{\rho} [ A_{\sigma}, A_{\lambda}
]^{K}  \\
 & & \mbox{ \hspace{2.0in} } + \: \frac{3}{5} A^{I}_{\mu} [ A_{\nu}, A_{\rho} ]^{J} [
A_{\sigma}, A_{\lambda} ]^{K} )  \: + \: 4 \mbox{ fermi } ] \\
\mbox{ } & & \mbox{ } \\
 & = \int_{M} [ \, & - \frac{1}{2} a_{IJ} D_{\mu} h^{I} D^{\mu} h^{J} \:
- \: \frac{1}{4} a_{IJ} F^{I}_{\mu
\nu} F^{J \mu \nu} \: - \: \frac{1}{2} a_{IJ} \overline{\lambda}^{i I}
\Gamma^{\mu} D_{\mu} \lambda^{J}_{i} \\ 
 & & - \: \frac{1}{2 \sqrt{6}} C_{IJK}  (
\overline{\lambda}^{i J} \Gamma^{\mu \nu} \lambda_{i}^{K} ) F^{I}_{\mu
\nu} \: -i \, a_{I J} \, \overline{\lambda}^{I i} \, [ h , \lambda_{i}
]^{J} \\
 & & + \: \frac{1}{ 6 \sqrt{6}} \epsilon^{\mu \nu \rho \sigma \lambda}
\, C_{IJK} \, ( (d A)^{I}_{\mu \nu} (d A)^{J}_{\rho \sigma}
A^{K}_{\lambda} \:
+ \frac{3}{2} (d A)^{I}_{\mu \nu} A^{J}_{\rho} [ A_{\sigma}, A_{\lambda}
]^{K}  \\
 & & \mbox{ \hspace{2.0in} }  + \: \frac{3}{5} A^{I}_{\mu} [ A_{\nu}, A_{\rho} ]^{J} [
A_{\sigma}, A_{\lambda} ]^{K} )  \: + \: 4 \mbox{ fermi } ]
\end{eqnarray*}

and its supersymmetry transformations are
\begin{eqnarray*}
\delta A^{I}_{\mu} & = & - \frac{1}{2} \, h^{I}_{a} \, ( 
\overline{\epsilon}^{i}
\Gamma_{\mu} \lambda^{a}_{i} ) \\
\delta \lambda^{a}_{i} & = & - \frac{i}{2} f^{a}_{x} \Gamma^{\mu}
D_{\mu} \phi^{x} \epsilon_{i} \: + \: \frac{1}{4} \Gamma^{\mu
\nu} \epsilon_{i} F^{I}_{\mu \nu}  h^{a}_{I} \\
\delta \phi^{x} & = & + \frac{i}{2} f^{x}_{a} ( \overline{\epsilon}^{i}
\lambda^{a}_{i} ) 
\end{eqnarray*}

with conventions
\begin{eqnarray*}
a_{IJ} & = & C_{IJ} \: + \: 2 i \sqrt{\frac{2}{3}} \, C_{IJK} \, h^{K} \\
C_{IJK} & \propto & \mbox{Tr } T_{I} \, \{ T_{J} , T_{K} \} \\
(dA)^{I}_{\mu \nu} & = & \partial_{\mu} A_{\nu}^{I} \: - \:
\partial_{\nu} A_{\mu}^{I} \\
F^{I}_{\mu \nu} & = & \partial_{\mu}A^{I}_{\nu} \, - \, \partial_{\nu}
A^{I}_{\mu} \, + \, [A_{\mu},A_{\nu}] \\
\partial_{y} f^{a}_{x} & = & \Gamma^{z}_{xy} \, f^{a}_{z} \: - \:
\Omega^{ab}_{y} \, f^{b}_{x} \\
h^{I}_{x} & = & \partial_{x} h^{I} \\
h^{I}_{a} & = & h^{I}_{x} \, f^{x}_{a} \\
h_{I}^{a} & = & a_{IJ} h^{J}_{a} \\
g_{xy} & = & a_{IJ} \, h^{I}_{x} \, h^{J}_{y} \\
\lambda^{I}_{i} & = & h^{I}_{a} \, \lambda^{a}_{i} 
\end{eqnarray*}

where $C_{IJ}$ and $C_{IJK}$ are constant and completely symmetric.
Again, we are closely following the conventions of
\cite{Sierra:matter,sugrav5Dmain,nonabel}.

Covariant derivatives are given by 
\begin{eqnarray*}
D_{\mu} \phi^{x} & = & \partial_{\mu} \phi^{x} \: - \: A_{\mu}^{I}
K_{I}^{x} \\
D_{\mu} h^{I} & = & \partial_{\mu} h^{I} \: + \: [A_{\mu},h]^{I} \\
D_{\mu} \lambda^{a}_{i} & = & \partial_{\mu} \lambda^{a}_{i} \: + \:
( D_{\mu} \phi^{x} ) \, \Omega^{ab}_{x} \, \lambda^{b}_{i} \: - \:
A^{I}_{\mu} \, L_{I}^{ab} \, \lambda_{i}^{b}  
\end{eqnarray*}

and gauge transformations are
\begin{eqnarray*}
\delta A^{I}_{\mu} & = & \partial_{\mu} \alpha^{I} \: + \: [ A_{\mu} ,
\alpha ]^{I} \\
\delta h^{I} & = & [ h , \alpha ]^{I} \\
\delta \phi^{x} & = & \alpha^{I} K^{x}_{I} \\
\delta \lambda^{a}_{i} & = & \alpha^{I} L^{ab}_{I} \lambda^{b}_{i}
\end{eqnarray*}

The $h^{I}$ are real adjoint-valued scalars, so under the action of
the nonabelian group they transform just as one would expect.  The
(equivalent) scalars $\phi^{x}$ transform nonlinearly under the
gauge group via the action of a Killing vector $K^{x}_{I}$.  
The Killing vectors form a representation of the gauge group:
\begin{displaymath}
\left[ \, K_{I} \, , \, K_{J} \, \right]^{x} \: = \: f^{K}_{IJ} \,
K^{x}_{K}
\end{displaymath} 
where $f^{K}_{IJ} \: = \: [ T_{I},T_{J} ]^{K}$.

By demanding self-consistency of these gauge transformations one can
derive several identities.  We will mention only two:
\begin{displaymath}
K^{x}_{J} \, h^{I}_{x} \: = \: h^{K} \, f^{I}_{KJ}
\end{displaymath}
\begin{displaymath}
K^{x}_{J} \, L^{a b}_{I, x} \: - \: K^{x}_{I} \, L^{a b}_{J, x} \: + \:
[  L_{I} , L_{J} ]^{a b} \: = \: f^{K}_{JI} \, L^{a b}_{K}
\end{displaymath}

Note that upon dimensional
reduction to four dimensions, this yields an N=2 theory with
prepotential of the general form
\begin{displaymath}
{\cal F} \: = \: C_{0} \: + \: C_{I} \, {\cal A}^{I}
 \: + \: \frac{1}{2} \, C_{IJ} \,
{\cal A}^{I} {\cal A}^{J} \: + \: \frac{1}{3!} C_{IJK} \, {\cal A}^{I} 
{\cal A}^{J} {\cal A}^{K}
\end{displaymath}

precisely as expected for very special geometry.

What about Fayet-Iliopoulos terms?  Under gauge transformations,
$D^{I}_{ij}$ transforms as
\begin{displaymath}
D^{I}_{ij} \: \rightarrow \: - \: [ \, \alpha \, , \, D_{ij} \,
]^{I}
\end{displaymath}
so if $D_{I ij}$ contains a constant term $r_{I ij}$ then $r_{I ij}$ can only 
be
nonzero when it lies in the center of the gauge group.

Suppose, for example, that the gauge group is $U(1)^{n}$, so that we can
add Fayet-Iliopoulos terms, in principle.  To proceed, simply add
the term 
\begin{displaymath}
- \: \frac{1}{4} \, a_{IJ} \, D^{I}_{ij} \, D^{J ij} 
\end{displaymath}
to the lagrangian, where
\begin{displaymath}
D^{I}_{ij} \: = \: a^{IJ} \, \left[ \, r_{J ij} \: - \: \sqrt{\frac{2}{3}} (
\overline{\lambda}^{a}_{i} \lambda^{b}_{j} ) h^{K}_{a} h^{L}_{b} C_{JKL}
\, \right]
\end{displaymath}
and $r_{I ij}$ is the Fayet-Iliopoulos Sp(1)-triplet of constants,  
and also add a term to $\delta \lambda^{a}_{i}$ :
\begin{displaymath}
\frac{1}{2} \,
D^{I}_{ij} \, h^{a}_{I} \, \epsilon^{j}
\end{displaymath}

We can couple the bulk theory to charged hyper plets by adding the terms
\begin{eqnarray*}
S_{bulk, hypers} & = \int_{M} [ \, & - \frac{1}{2} D_{\mu} \sigma^{ia'}
\, D^{\mu} \sigma_{ia'} \: - \: \frac{1}{4} \overline{\lambda}^{a'}
\Gamma^{\mu} D_{\mu} \lambda_{a'} \: - \: \frac{1}{2} F^{ia'} F_{ia'} \:
- \: \frac{1}{4} a_{IJ} \, D^{I}_{ij} D^{J ij} \\
 & & + i \, \sigma^{ia'} h^{I}_{a} (T_{I})_{a'b'}
(\overline{\lambda}^{b'} \lambda^{a}_{i} ) \: + \: \frac{i}{4} (
\overline{\lambda}_{a'} \lambda_{b'} ) h^{I} (T_{I})^{a'b'} \: + \: 4
\mbox{ fermi } ]
\end{eqnarray*}

with auxiliary fields,
\begin{eqnarray*}
F^{ia'} & = & h^{I} (T_{I})^{a'}_{\: \: b'} \, \sigma^{ib'} \\
D^{I}_{ij} & = & a^{IJ} \left[ r_{J ij} \: + \: \sigma_{ia'} (T_{J})^{a'b'} 
\sigma_{jb'}  \:
- \: \sqrt{\frac{2}{3}}  ( \overline{\lambda}^{a}_{i}
\lambda^{b}_{j} ) h^{K}_{a} h^{L}_{b} C_{JKL} \right] 
\end{eqnarray*}

gauge transformations
\begin{eqnarray*}
\delta \sigma^{i a'} & = & \alpha^{I} \, (T_{I})^{a'}_{\: \: \: b'} \,
\sigma^{i b'} \\
\delta \lambda^{a'} & = & \alpha^{I} \, (T_{I})^{a'}_{\: \: \: b'} \,
\lambda^{b'}
\end{eqnarray*}

and covariant derivatives
\begin{eqnarray*}
D_{\mu} \sigma^{i a'} & = & \partial_{\mu} \sigma^{i a'} \: - \:
A^{I}_{\mu} \, (T_{I})^{a'}_{\: \: b'} \, \sigma^{i b'} \\
D_{\mu} \lambda^{a'} & = & \partial_{\mu} \lambda^{a'} \: - \:
A^{I}_{\mu} \, (T_{I})^{a'}_{\: \: b'} \, \lambda^{b'}
\end{eqnarray*}

Note that the hyper plet moduli space is assumed flat, in order
to considerably simplify notation.  For more general treatments see
\cite{hitchin}.

The $(T_{I})^{a'b'}$ are constant matrices giving the action of the gauge
group on the hyper plets.  We've assumed the gauge group acts
only on Sp(n) indices, so that, for example, the supersymmetry
transformation parameters are neutral under the gauge group.  
In conventions used here, 
\begin{displaymath}
(T_{I})^{a' b'} \, (T_{J})_{b'}^{\: \: \: c'} \: - \: (T_{J})^{a' c'} \,
(T_{I})_{b'}^{\: \: \: c'} \: = \: f^{K}_{IJ} \, (T_{K})^{a' c'}
\end{displaymath}
\begin{displaymath}
(T_{I})^{a' b'} \: = \: + \, (T_{I})^{b' a'}
\end{displaymath}

The supersymmetry transformations are
\begin{eqnarray*}
\delta A^{I}_{\mu} & = & - \frac{1}{2} h^{I}_{a} (
\overline{\epsilon}^{i}
\Gamma_{\mu} \lambda^{a}_{i} ) \\
\delta \lambda^{a}_{i} & = & - \frac{i}{2} f^{a}_{x} \Gamma^{\mu}
( D_{\mu} \phi^{x} ) \epsilon_{i} \: + \: \frac{1}{4} \Gamma^{\mu
\nu}
\epsilon_{i} F^{I}_{\mu \nu} h^{a}_{I} \: + \: \frac{1}{2} D^{I}_{ij}
h^{a}_{I} \epsilon^{j} \: + \: 3 \mbox{ fermi } \\
\delta \phi^{x} & = & + \frac{i}{2} f^{x}_{a} ( \overline{\epsilon}^{i}
\lambda^{a}_{i} ) \\
\delta \sigma^{ia'} & = & \frac{i}{2} ( \overline{\epsilon}^{i}
\lambda^{a'} ) \\
\delta \lambda^{a'} & = & - i \Gamma^{\mu} ( D_{\mu} \sigma^{ia'} )
\epsilon_{i} \: + \: F^{ia'} \epsilon_{i} \: + \: 3 \mbox{ fermi } \\
\end{eqnarray*}

Ordinarily [in 4D N=1, for example] one expects that the
Fayet-Iliopoulos term can be shifted by quantum corrections
[proportional to the sum of the charges of the chiral plets in 4D N=1].
Here, however, it is easy to show that the renormalization of the 
Fayet-Iliopoulos term is
\begin{displaymath}
\delta D_{I}^{ij} \: \propto \: \epsilon^{ij} \, (T_{I})^{a' b'} \,
\Omega_{a' b'} \, \int \frac{ d^{5} k }{k^{2}} \: = \: 0 
\end{displaymath}

so the shift vanishes by symmetry.  In retrospect this is not
surprising:  consider 4D N=2 QED.  Each hyper plet is composed of a pair
of chiral plets, with equal and opposite charges, so that the sum of the
charges of the chiral plets always vanishes, so the N=1 D term does
not get shifted by quantum corrections.

Note that at a generic point in the low energy theory, the gauge group
action on the hyper plets will be Higgsed away, so 
the hyper plets will live on a hyperKahler reduction
implicit in the above, with hyperKahler moment maps the 
D term triplets $D^{I}_{ij}$.  For example, we can construct
$A_{n}$ surface singularities as classical Higgs moduli spaces, following
\cite{kronheimer}.  Such a theory has gauge group $U(1)^{n}$, and n+1
charged hyper plets, the ith hyper plet charged under the ith and
(i+1)th $U(1)$s.  The Fayet-Iliopoulos parameters $r_{I ij}$ are
the Kahler classes of the exceptional divisors with respect to
each of the three complex structures.  Precisely this situation arises
in type II compactifications near conifolds \cite{conifold,gmv}.
Consider for example type IIB near a conifold singularity at which some
number of $S^{3}$s are collapsing, then as shown in \cite{gmv} in the
low-energy effective field theory the hyper plet moduli space can be
locally approximated as having an $A_{n}$ singularity.

How do we orientifold this theory?  As suggested in section two,
there are two possible orientifolds of a theory with bulk vectors.  One 
preserves
vectors on the boundary, the other projects the bulk vector
plets to boundary chiral plets.  
We will first consider orientifolding this theory so as to preserve vectors on
the boundary.  The orientifold symmetry is given by 
\begin{eqnarray*}
\phi^{x} & \rightarrow & - \, \phi^{x} \\
h^{I} & \rightarrow & - \, h^{I} \\
A^{I}_{\mu} & \rightarrow & + \, A^{I}_{\mu}  \mbox{ for } \mu < 5  \\
C_{IJ} & \rightarrow & + \, C_{IJ} \\
\lambda^{ia} & \rightarrow & + i \, \Gamma_{5} \, \lambda^{a}_{i} \\
\lambda^{a}_{i} & \rightarrow & - i \, \Gamma_{5} \, \lambda^{ia} \\
f^{a}_{x} & \rightarrow & - \, f^{a}_{x} \\
\sigma^{ia'} & \rightarrow & \sigma_{ia'} \\
\lambda^{a'} & \rightarrow & + i \, \Gamma_{5} \, \lambda_{a'} \\
\lambda_{a'} & \rightarrow & - i \, \Gamma_{5} \, \lambda^{a'} \\
(T_{I})^{a'b'} & \rightarrow & (T_{I})_{a'b'} \\
D^{I}_{ij} & \rightarrow &  + \, D^{I ij} \\
F^{ia'} & \rightarrow & - \, F_{ia'}
\end{eqnarray*}

and note this can only be a symmetry of the action when $C_{IJK} \: = \:
0$.  Recall from section two that 5D vector plets project to boundary
vectors consistently only on subsets of extended Kahler moduli space, 
precisely as is clear here from constraints on the
intersection numbers \cite{anton} $C_{I J K}$ in the 
low-energy effective action.  

The supersymmetries that commute with the orientifold are given by
\begin{eqnarray*}
\epsilon_{i} & = & + i \, \Gamma_{5} \, \epsilon^{i} \\
\epsilon^{i} & = & - i \, \Gamma_{5} \, \epsilon_{i}
\end{eqnarray*}

We can put a superpotential on the boundary of this theory in almost
the same fashion as previously.
First, add a term to the supersymmetry transformation of $\lambda^{a'}$:
$ \delta(x_{5}) \, G^{ia'} \, \epsilon_{i} $,
where in the notation of the last section
\begin{eqnarray*}
G^{1 A} & = & \frac{1}{2} \, [ \, \partial_{A}
W \: + \: \partial^{*}_{A} \overline{W} \, ] \\
G^{2 A} & = & \frac{1}{2i} \, [ \,
\partial_{A} W \: - \: \partial^{*}_{A}
\overline{W} \, ]
\end{eqnarray*}

The superpotential is now a gauge-invariant holomorphic function
of the boundary chiral superfields.
The fields of the
boundary chiral plet are
\begin{eqnarray*}
A^{A} & = & \sigma^{ 1 A } \: + \: i \,
\sigma^{ 2 A } \\
\chi^{ A } & = & \lambda^{ A } \: - \: i \,
\lambda^{ A' }
\end{eqnarray*}

By restricting the bulk supersymmetry transformations to the boundary we
find
\begin{eqnarray*}
\delta A^{ A } & = & \frac{i}{4} \, \overline{\epsilon}
\chi^{ A } \\
\delta A^{ A * } & = & \frac{i}{4} \,
\overline{\epsilon}^{*} \chi^{ A * } \\
\delta \chi^{ A } & = & - \, \Gamma^{\mu} \, ( D_{\mu}
A^{ A } ) \, \epsilon^{*} \: + \: i \, \delta(0) \, (
\partial^{*}_{ A } \overline{W} ) \, \epsilon \: + \: 3
\mbox{ fermi} \\
\delta \chi^{ A * } & = & + \, \Gamma^{\mu} \, ( D_{\mu}
A^{ A * } ) \, \epsilon \: - \: i \, \delta(0) \, (
\partial_{ A } W ) \, \epsilon^{*} \: + \: 3 \mbox{ fermi}
\end{eqnarray*}

with the boundary interaction

\begin{displaymath}
S_{boundary} \: = \: \int_{ \partial M } \, [ \, ( \partial_{
A } W ) \, ( \partial^{*}_{ A } \overline{W} )
\, \delta(0) \: - \: \frac{1}{8} \, ( \partial_{ A }
\partial_{ B } W )  \, \overline{\chi}^{ A }
\chi^{ B }  \: + \: \frac{1}{8} ( \partial^{*}_{
A } \partial^{*}_{ B } \overline{W} ) \,
\overline{\chi}^{ A * } \chi^{ B * } \, ] 
\end{displaymath}

There is a second orientifold of this theory, which projects the bulk
vector plets to boundary chiral plets.
The relevant orientifold symmetry is
\begin{eqnarray*}
\phi^{x} & \rightarrow & + \, \phi^{x} \\
h^{I} & \rightarrow & + \, h^{I} \\
A_{\mu}^{I} & \rightarrow & - \, A_{\mu}^{I} \mbox{ for } \mu  <  5 \\
C_{IJ} & \rightarrow & + \, C_{IJ} \\
C_{IJK} & \rightarrow & + \, C_{IJK} \\
\lambda^{ia} & \rightarrow & + i \Gamma_{5} \lambda_{i}^{a} \\
\lambda_{i}^{a} & \rightarrow & - i \Gamma_{5} \lambda^{ia} \\
f_{x}^{a} & \rightarrow & + \, f_{x}^{a} 
\end{eqnarray*}

and similarly for the hyper plets.  The supersymmetries that commute
with this orientifold are also as before.

Note that chiral plets obtained from bulk vector plets in this manner can not 
couple
perturbatively to a superpotential:  perturbatively, the 5th component
of a $U(1)$ vector $A_{5}^{I}$ has
a gauge symmetry $A_{5}^{I} \: \rightarrow \: A_{5}^{I} \: + \:
\partial_{5} \epsilon$, so $A_{5}^{I}$ behaves like an axion on the
boundary.  However, this Peccei-Quinn-like symmetry can be broken by 
nonperturbative effects.
For example, in the M theoretic description of the heterotic string,
worldsheet instantons are simply 2-branes stretched between the two ends
of the world, and these 2-branes will break the gauge symmetry.  In
fact, this is precisely the M-theoretic understanding of the old
sigma model nonrenormalization theorem for the spacetime superpotential 
\cite{xw}.

One might be curious about gauge-invariance of the Chern-Simons term
in a boundary theory.  The boundary terms that
are generated upon gauge-transformation of the Chern-Simons term all vanish on
the boundary\footnote{Note this is in contrast to the usual situation
in 3D.  Given a 3D Chern-Simons form $I_{CS}$ on a three-manifold M
with boundary, $e^{i I_{CS}}$ is not well-defined under gauge
transformations but rather picks up a factor due to boundary terms, and
so is interpreted as a section of a bundle over the space of connections
on $\partial M$.  Were it not for the orientifold boundary conditions
an analogous phenomenon would occur in the 5D theory being discussed.},
because of the boundary condition on the gauge parameter,
leaving only the usual $\int_{M} ( g^{-1} dg )^5$ term,
which fixes the coefficient of the Chern-Simons term to be proportional
to an integer.

{\bf 5.  Anomalies}

In this section, we will explain the strong coupling version of
a familiar fact about the weakly coupled heterotic string.
In heterotic string compactification on a six-manifold $X$ 
with some  gauge sheaf $V$, one often obtains
at tree level what appears to be -- just going by the massless
fermion spectrum -- an anomalous $U(1)$ \cite{4Danom,dg}. Let $A^{(1)}_{\mu}$ be the 
gauge field with apparently anomalous couplings, 
and let $F=dA$. In this situation, if one looks closely, one finds always
a scalar field $a$ with a coupling $\partial_{\mu} a \, A^{(1) \mu}$
and a gauge transformation law just right to cancel the anomaly.
All this is ensured by the Green-Schwarz anomaly cancellation
mechanism in ten dimensions.  The physical consequence is that $A^{(1)}_{\mu}$
becomes massive by Higgsing of $a$.   We want to see how this
scenario works out in the strongly coupled region of the $E_8\times
E_8$ heterotic string, related to physics on $R^4\times S^1/Z_2$.

How are anomalies cancelled in models seen as 5D orientifolds?
The chiral anomalies
arising on the 4D boundary of the 5D effective theory are cancelled by
bulk 5D Chern-Simons terms that descend from the 11D 3-form potential bosonic
interaction, and the $U(1)$ itself is Higgsed by terms descending from
Green-Schwarz interactions.

Before demonstrating this more precisely, we will review the Green-Schwarz
mechanism in 
M theory from \cite{pwII}.  
The variation of
the boundary effective action after gauge transformation of the boundary
$E_{8}$ vector is proportional to
\begin{displaymath}
\int_{M^{10}} \epsilon^{M_{1} M_{2} \ldots M_{10}} \, tr( \epsilon
F_{M_{1} M_{2}} ) \, tr( F_{M_{3} M_{4}} F_{M_{5} M_{6}} ) \,
tr ( F_{M_{7} M_{8} } F_{M_{9} M_{10}} )
\end{displaymath}
This anomaly is cancelled by the variation of a Chern-Simons-like term
in the 11D bulk.  Specifically, 
recall the bosonic 3-form potential
has the 11D coupling
\begin{displaymath}
\int_{M^{11}} \left[ \, - \:
\frac{\sqrt{2}}{3456} \, \epsilon^{I_{1} I_{2} \ldots I_{11}} \, C_{I_{1}
I_{2} I_{3}} \, G_{I_{4} \ldots I_{7}} \, G_{I_{8} \ldots I_{11}}
\right]
\end{displaymath}
and also recall gauge transformations of the boundary vector are
accompanied by a gauge transformation of the three-form potential
\begin{displaymath}
\delta C_{11 A B} \: = \: - \frac{\kappa^{2}}{6 \sqrt{2} \lambda^{2}}
\delta (x^{11}) tr \epsilon F_{A B}
\end{displaymath}
so under a gauge transformation of the boundary vector the 11D
Chern-Simons-like interaction picks up the variation 
\begin{displaymath}
\int_{M^{10}} \epsilon^{M_{1} M_{2} \ldots M_{10}} \, tr( \epsilon F_{
M_{1} M_{2}} ) \, G_{M_{3} \ldots M_{6}} \, G_{M_{7} \ldots M_{10} }
\end{displaymath}
Finally, recall because of the Bianchi identity we have the result
\begin{displaymath}
G_{A B C D} \: = \: - \frac{3}{\sqrt{2}} \frac{\kappa^{2}}{\lambda^{2}}
\epsilon (x^{11}) F^{a}_{[ A B} F^{a}_{ C D ]} \: + \: \cdots
\end{displaymath}
(ignoring Riemann curvature terms) so the variation of the
11D Chern-Simons-like term is proportional to the
variation of the effective action due to the 10D chiral anomaly,
and so we cancel the anomaly.

What happens after compactification?  To be specific, consider embedding
a gauge sheaf $V$ of the form $E \oplus L$ in one of the $E_{8}$s, where
$c_{1}(E) \: = \: - \, c_{1}(L)$ and $L$ is rank 1, $E$ is rank 4, then
$E_{8}$ is broken to $SU(5) \times U(1)$.  In particular, consider the 
$U(1)^{3}$ anomaly in 4D.  Let $F_{ i \overline{\jmath}}$ be the de Rham
image of 
$c_{1}(L)$.  Then the 4D axion obtained from compactifying $C_{11 A B}$
on $F_{i \overline{\jmath}}$ transforms under gauge transformations of the
low-energy $U(1)$.  If we denote by $A_{\mu}^{(1)}$ the
low-energy boundary $U(1)$, then because of the $U(1)^{3}$ anomaly under
gauge transformations we pick up a contribution to the boundary
effective action
\begin{displaymath}
\int \, \epsilon^{\mu \nu \rho \sigma} \, \epsilon \, F^{(1)}_{\mu \nu}
\, F^{(1)}_{\rho \sigma}
\end{displaymath}
The 5D bulk theory contains a Chern-Simons term of the form
\begin{displaymath}
\int \, \epsilon^{\mu \nu \rho \sigma \delta} \, A_{\mu}^{I} \, F_{\nu 
\rho}^{J}
\, F_{\sigma \delta}^{K}
\end{displaymath}
where each 5D vector is obtained by compactifying the 11D 3-form
potential.  Because gauge variations of 10D vectors are coupled to gauge
variations of the 11D 3-form potential, we can read off that gauge
variations of 4D boundary vectors are coupled to gauge variations of
these 5D vectors.  Thus, under a gauge transformation of
$A_{\mu}^{(1)}$, the 5D Chern-Simons term transforms as
\begin{displaymath}
\int \, \epsilon^{\mu \nu \rho \sigma 5} \, \epsilon \,F^{J}_{\mu \nu} \,
F^{K}_{\rho \sigma}
\end{displaymath}
and by compactifying $G_{A B C D}$ on $F_{i \overline{\jmath}}$ to get the 
remaining 5D vectors, we can read off the 5D boundary conditions from
the 11D boundary conditions on $G_{A B C D}$, to find that the variation
of the 5D Chern-Simons term under a gauge variation of $A_{\mu}^{(1)}$
is proportional to
\begin{displaymath} 
\int \, \epsilon^{\mu \nu \rho \sigma 5} \, \epsilon \, F^{(1)}_{\mu
\nu} \, F^{(1)}_{\rho \sigma} 
\end{displaymath}
which is exactly as needed to cancel out the 4D boundary chiral anomaly.

By compactifying the $G_{A B C D}$ on other field strengths,
we can cancel out more general nonabelian anomalies and even
gauge-gravitational-gravitational anomalies (by using the Riemann
curvature terms that have been suppressed so far).  For example, by
compactifying one of the 11D $G_{A B C D}$ factors on $c_{2}(E)$, then
the boundary condition on the other $G_{A B C D}$ factor makes it
proportional to $tr F \wedge F$, where the trace runs over both $U(1)$ and
$SU(5)$ indices, thereby yielding a contribution to both the $U(1)^{3}$
and $U(1)-SU(5)^{2}$ anomalies. 

In the analysis sketched above, the $C_{I J K}$ factor in the 5D
Chern-Simons term was suppressed.  As noted in \cite{anton}, this factor
is
proportional to the obvious intersection form on the 3-fold.  For
example, the first case studied (which contributes to $U(1)^{3}$) would
have $C_{I J K}$ proportional to $< X \, | \, c_{1}(E)^{3} >$.  More to
the point, for the anomaly cancellation mechanism outlined above to
function, it must be the case that all contributions to the anomaly
factorize, ie, must be of the form $< X \, | \, c_{1}(E) \cup \cdots
>$.  This is precisely the compactification of the 10D Green-Schwarz
factorization condition.

Let's work through the $SU(5) \times U(1)$ example in detail, following
\cite{dg}.  As our gauge sheaf $V$ has rank 5, we can expect ${\bf 10}$s of
$SU(5)$ corresponding to elements of $H^{m}(X,V)$, ${\bf \overline{5}}$s of
$SU(5)$ corresponding to elements of $H^{m}(X,\wedge^{2} V)$, and some $SU(5)$
singlets corresponding to elements of $H^{m}(X, \mbox{End } V)$.  Let
the $U(1)$ subgroup of the structure group\footnote{For simplicity,
we are assuming $E$, $L$ are well-defined bundles.} that yields the low-energy
$U(1)$ be defined by $\mbox{diag}(1,1,1,1,-4)$, then we have the
following chiral plets charged under the low-energy $U(1)$:
\begin{center}
\begin{tabular}{|l|l|c|} \hline
$SU(5)$ representation & Sheaf cohomology group & $U(1)$ charge \\
\hline
${\bf 10}$ & $H^{m}(X , E)$ & 1 \\
     & $H^{m}(X , L)$ & -4 \\ \hline
${\bf \overline{5}}$ & $H^{m}(X , \wedge^{2} E)$ & 2 \\
               & $H^{m}(X , E \otimes L)$ & -3 \\ \hline
Singlets       & $H^{m}(X , E \otimes L^{-1} )$ & 5 \\ \hline
\end{tabular}
\end{center}
Given this information we can rapidly compute some anomalies.
For example, the $U(1)$ trace anomaly is
\begin{displaymath}
(10) \left[ \, \chi(E) \: - \: 4 \, \chi(L) \, \right] \: + \: (5)
\left[ \, 2 \, \chi(\wedge^{2} E) \: - \: 3 \, \chi(E \otimes L) \,
\right] \: + \: \left[ \, 5 \, \chi(E \otimes L^{-1}) \, \right]
\end{displaymath}
where 
\begin{displaymath}
\chi (V) \: = \: \sum_{i} \, (-)^{i} \, \mbox{ dim } H^{i} ( X , V )
\end{displaymath}
Using the Hirzebruch-Riemann-Roch theorem \cite{hirzebruch}
\begin{displaymath}
\chi (V) \: = \: < X \, | \, ch(V) \cup td(TX) >
\end{displaymath}
we can rewrite the $U(1)$ trace anomaly as
\begin{displaymath}
< X \, | \, c_{1}(E) \, \cup \, \left[ \, \frac{65}{3} \, c_{1}(E)^{2} \: - \: 
30 \, c_{2}(E)
\: + \: \frac{15}{2} \, c_{2}(TX) \, \right] >
\end{displaymath}
This $U(1)$ trace anomaly is proportional to the
$U(1)-\mbox{graviton}^{2}$ triangle diagram.  If it is nonzero, then we
get a Fayet-Iliopoulos term generated at 1-loop.  Such a term would
spontaneously break supersymmetry and thereby destabilize the vacuum in
a Coulomb phase --
but as will be shown later, the $U(1)$ is Higgsed, so the
Fayet-Iliopoulos term merely shifts some scalar vevs, rather than
breaking supersymmetry.

Proceeding similarly, the $U(1)^{3}$ anomaly is 
\begin{eqnarray*}
\mbox{ } & & (10) \left[ \, \chi(E) \: - \: (64) \, \chi(L) \right] \: +
\: (5) \left[ \, (8) \, \chi(\wedge^{2} E) \: - \: (27) \, \chi(E
\otimes L) \, \right] \: + \: \left[ \, (125) \, \chi(E \otimes L^{-1})
\, \right] \\
\mbox{ } & & = \: < X \, | \, c_{1}(E) \, \cup \, \left[ \,
\frac{770}{3} \, c_{1}(E)^{2} \: - \: 300 \, c_{2}(E) \: + \: 85 \, c_{2}(TX) 
\,
\right] >
\end{eqnarray*}
and the $U(1) - SU(5)^{2}$ anomaly is
\begin{eqnarray*}
\mbox{ } & & \left[ \, \chi(E) \: - \: 4 \, \chi(L) \right] \,
d^{(2)}({\bf 10}) \: + \: \left[ \, (2) \, \chi(\wedge^{2} E) \: - \: 3
\, \chi(E \otimes L) \, \right] \, d^{(2)}({\bf \overline{5}}) \\
\mbox{ } & & = \: < X \, | \, c_{1}(E) \, \cup \, \left[ \, \frac{9}{2}
\, c_{1}(E)^{2} \: - \: 5 \, c_{2}(E) \: + \: \frac{7}{4} \, c_{2}(TX)
\, \right] > \, d^{(2)}({\bf \overline{5}})
\end{eqnarray*}
using \cite{mckay} $d^{(2)}({\bf 10}) \: = \: 3 \, d^{(2)}({\bf
\overline{5}})$.  As anticipated earlier, in each case the $< X \, | \,
c_{3}(E) \, >$ contribution cancels, so the anomaly factorizes.

In addition we can also easily see how anomalous $U(1)$s are Higgsed,
closely following \cite{4Danom,dg}.  Let's first review how this works
in standard heterotic compactifications.  Recall that the torsion picks
up a couple Chern-Simons terms, so in 10D the kinetic term for the
antisymmetric tensor has the form
\begin{displaymath}
(  H_{A B C} \, + \, \omega(YM)_{A B C} \, + \, \omega(R)_{A B C}  )^{2}
\end{displaymath}
By compactifying on $F_{i \overline{\jmath}}$ (and ignoring Riemann
curvature terms) one component of this becomes
\begin{displaymath}
< F >^{2} \, ( \partial_{\mu} a \, + \, A_{\mu}^{(1)} )^{2}
\end{displaymath}
(Note if we did not know in advance that the axion $a$ had a translation
symmetry gauged by $A_{\mu}^{(1)}$, we could have deduced it from the
coupling above.)  This has precisely the effect of Higgsing the $U(1)$,
as desired.

Now, how does this work in M theory compactifications?  Recall that the 11D 
low-energy action has a kinetic term for the 3-form
potential proportional to $G_{I J K L} G^{I J K L}$.  In particular,
consider the $G_{11 A B C} G^{11 A B C}$ component.  Recall from
\cite{pwII} that the modified Bianchi identity is solved by modifying
$G_{I J K L}$ as
\begin{displaymath}
G_{11 A B C} \: = \: ( \partial_{11} C_{A B C} \: \pm \: \cdots ) \: +
\: \frac{\kappa^{2}}{\sqrt{2} \lambda^{2}} \, \delta(x^{11}) \, \omega_{A B
C}
\end{displaymath}
(ignoring Riemann curvature terms, once again)
so following the $SU(5) \times U(1)$ example above, there is a 
$G_{11 \mu i \overline{\jmath}} G^{11 \mu i \overline{\jmath}}$ term 
which is of the
form 
\begin{displaymath}
 < F >^{2} \, ( F_{(11) \mu} \, + \, \delta(x^{11}) A_{\mu}^{(1)})^{2}
\end{displaymath}
where $F_{(11) \mu}$ is the field strength of the 5D $U(1)$ that
descends from $G_{I J K L}$.  This $U(1)$ vector projects to a
chiral plet on the boundary, the chiral plet containing the axion $a$
mentioned in the last paragraph.  Again, $A_{\mu}^{(1)}$ gauges a
translation symmetry of $a$, effectively, and the boundary-confined $U(1)$ is
Higgsed.  World-sheet instantons (membranes stretched between the ends
of the world) break the remaining global symmetry. 

Naively this would appear to Higgs any $U(1)$; why does it 
only Higgs anomalous $U(1)$s?  This is implicit in the existence of a
cohomologically nontrivial field strength $F_{i \overline{\jmath}}$ on which
to compactify the 11D 3-form potential.  If it had been the case that
$c_{1}(L) \: = \: 0$, then $C_{11 i \overline{\jmath}}$ would have been
gauge-trivial, and so we would have gotten neither an axion $a$ nor an 
interaction term $\delta(x^{11}) \, A_{\mu}^{(1)}$ in the 4D theory,
so the $U(1)$ vector $A_{\mu}^{(1)}$ would not have been Higgsed.  
In such a case the $U(1)$ is not anomalous, so there is no
difficulty. 

So far we have considered anomalies due (primarily) to boundary-confined
vectors.  What about boundary vectors that descend from bulk vector
plets?  The boundary conditions for such an orientifold
demand that $C_{IJK} \: = \: 0$, so if there are any boundary chiral
anomalies in such vectors, they can not be cancelled by a bulk
Chern-Simons term.

{\bf 6.  Discussion}

In this paper we have worked out explicit lagrangians describing
superpotential coupling to the boundary of a 5D orientifold, as relevant
to a number of compactifications, and also made some general comments on
compactifications of 11D M theory orientifolds, relevant to the strong
coupling limit of the $E_{8} \times E_{8}$ heterotic string.

Note that the superpotentials we have discussed in this paper do not
include couplings generated by membranes stretched between the ends
of the world (worldsheet instantons in standard heterotic
compactifications).  Such objects can not be described locally in five
dimensions.  When they are absent (or, when the radius of the fifth
dimension is large, so that they are exponentially suppressed), we have  
a nonrenormalization theorem specifying that chiral plets descending
from bulk vectors do not couple to a superpotential.  In heterotic
string compactifications this is a well-known result \cite{xw},
but note that this also gives constraints on superpotentials of the models
of types (2) and (3) discussed in section two.

Note also that in orientifolds there are new ways to spontaneously break
supersymmetry.
For example, put some simple
O'Raifeartaigh model on a boundary, with chiral plets all descending
from bulk vector plets.  Then we have spontaneously broken supersymmetry
due to a purely boundary interaction !  Essentially the same idea has
been discussed in \cite{pIII}.

There are additional ways to break supersymmetry in these models.
Consider a 5D theory with a single hyper plet.  Put a
superpotential on each boundary of the form $\lambda \Phi \: + \: m
\Phi^{2}$, with distinct $\lambda$, $m$ on each boundary.  This
superpotential uniquely fixes a nonzero vev for the boundary chiral
plet, but as the couplings are different on the boundaries, the vevs are
distinct on the boundaries.  It is easy to see that this
spontaneously breaks supersymmetry in the 5D bulk.  Unlike the case
above where
supersymmetry was clearly broken locally on the boundary, here
supersymmetry is naively unbroken on the boundary, and is only broken by
the global 5D topology.

Moreover, in this strong-coupling description of the heterotic string,
it may be possible to derive certain heterotic nonperturbative effects.  For example,
consider
compactifying M theory on a singular Calabi-Yau.  Suppose for
definiteness the Calabi-Yau contains a genus g curve of $A_{n}$
singularities, then following \cite{kmp} we should expect to recover
SU(n+1) gauge theory with g adjoint hyper plets.  [Recall this is the
classical result, but following \cite{nikita} Seiberg-Witten theory in
5D is trivial.]
The orientifold action that yields the $E_{8} \times
E_{8}$ heterotic string will then project each vector plet to a boundary
chiral plet, so in the low energy theory it seems likely that we will find g+1
chiral plets that are in the adjoint representation of a global symmetry
group, SU(n+1) (g from bulk hyper plets, 1 from bulk vector plet).  In
addition,
as the base space is singular there are no
doubt small instanton effects that also need to be taken into account.
Thus, ignoring small instanton effects we see that compactifying
heterotic $E_{8} \times E_{8}$ on a singular Calabi-Yau 3-fold of this
form may yield
new massless neutral matter and an enhanced global symmetry group, the
projection of bulk 5D vector plets.\footnote{Also note that this means
the ``end-of-the-world" branes in M theory may have
many properties expected of brane probes \cite{braneprobe} such as
possessing enhanced global symmetry in the presence of background
enhanced local symmetry.}  Work on this approach to studying
nonperturbative effects in heterotic string theory is in progress.

Finally, we will note that this technology may have broader
applications than have been discussed in this paper.  One such possible
application is a generalization of Seiberg duality away from the IR
limit.  A 5D orientifold such as has been discussed here looks like a 4D
N=1 theory at very low energies, where all wavelengths are much longer
than the size of the 5th dimension.  As one strong coupling limit of the
4D compactification of the $E_{8} \times E_{8}$ heterotic string is five
dimensional, perhaps some insight into Seiberg duality can be gained
by going to five dimensions.

{\bf Acknowledgements}

We would like to thank P. Horava and especially E. Witten for 
useful conversations.

{\bf Appendix on Conventions}

The metric $\eta \: = \: \mbox{diag}(-,+,+,+,+)$

The index $i$ is in the fundamental of Sp(1),
the index $a'$ is in the fundamental of Sp(n).

Sp(1) indices are raised and lowered as
\begin{eqnarray*}
V^{i} & = & \epsilon^{ij} \, V_{j} \\
V_{i} & = & V^{j} \, \epsilon_{ji}
\end{eqnarray*}

with $\epsilon_{12} \: = \: \epsilon^{12} \: = 1$, so for example $V_{i}
\, W^{i} \: = \: - \, V^{i} \, W_{i}$.
The Sp(n) indices behave identically, under the action of the
Sp(n)-invariant tensor $\Omega_{a'b'}$. 

Note that the Sp(n) spin connection has the amusing property
\begin{displaymath}
\omega_{x'}^{a' b'} \: = \: + \, \omega_{x'}^{b' a'}
\end{displaymath}

that is, it is symmetric rather than antisymmetric in its ``local
Lorentz" indices, because the adjoint representation of Sp(n) is a symmetric tensor
in its fundamental representations, as opposed to SO(n).

All 5D spinors are symplectic-Majorana.  Boundary 4D spinors are, by
construction, Weyl.
Five-dimensional spinor conventions are such that
\begin{displaymath}
\overline{\psi}^{i} \, \Gamma^{\nu_{1} \cdots \nu_{n}} \epsilon^{j} \:
= \: + \: \overline{\epsilon}^{j} \Gamma^{\nu_{n} \cdots \nu_{1} }
\psi^{i}
\end{displaymath}

Gamma matrices are defined in ``strength one" conventions:
\begin{displaymath}
\Gamma^{\nu_{1} \cdots \nu_{n}} \: = \: \frac{1}{n!} \left[
\Gamma^{\nu_{1}} \Gamma^{\nu_{2}} \cdots \Gamma^{\nu_{n}} \: \pm \: 
\cdots \: \right]
\end{displaymath}

\end{document}